\documentclass[11pt]{article}
\usepackage{latexsym}
\usepackage{amsmath,amssymb,color}
\usepackage{graphicx}

\usepackage{bbold}

\DeclareMathAlphabet{\mathpzc}{OT1}{pzc}{m}{it}

\newcommand{\ba}{\begin{eqnarray}}
\newcommand{\ea}{\end{eqnarray}}
\newcommand{\be}{\begin{equation}}
\newcommand{\ee}{\end{equation}}

\def\3s{{s \choose 3}}
\def\4s{{s \choose 4}}
\def\5s{{s \choose 5}}
\def\6s{{s \choose 6}}
\def\12{\frac{1}{2}}

\def\bec{\begin{center}}
\def\ec{\end{center}}

\def\cD{{\cal D}}

\def\tr{{\rm tr}}

 \def\det{{\rm det\,}}
\def\be{\begin{equation}}
\def\ee{\end{equation}}
\def\bea{\begin{eqnarray}}
\def\eea{\end{eqnarray}}
\def\ba{\begin{array}}
\def\ea{\end{array}}

\begin{document}

\title{Translation invariant time-dependent solutions to massive gravity}

\author{J.~Mourad\footnote{mourad@apc.univ-paris7.fr} $\;$and D.A.~Steer\footnote{steer@apc.univ-paris7.fr}\\
{\it AstroParticule \& Cosmologie,}\\
{\it UMR 7164-CNRS, Universit\'e Denis Diderot-Paris 7,}\\
{\it CEA, Observatoire de Paris, F-75205 Paris Cedex 13, France}}
\maketitle

{\bf Abstract}

{ Homogeneous time-dependent solutions of massive gravity generalise the plane wave solutions of the linearised Fierz-Pauli equations for a massive spin-two particle, as well as the Kasner solutions of General Relativity.  We show that they also allow a clear 
counting of the degrees of freedom and represent a simplified  framework to work out the constraints,
  the equations of motion and the initial value formulation. We work in the vielbein formulation of massive gravity, find the phase space resulting from the constraints and show that several disconnected sectors of solutions exist some of which are unstable. The initial values determine the sector to which a solution belongs. 
  Classically, the theory is not pathological but quantum mechanically the theory may suffer from instabilities. The latter are not due to an extra ghost-like degree of freedom. }

\tableofcontents
\section{Introduction}

The study of possible deformations of general relativity is of major theoretical and phenomenological importance \cite{Rubakov:2008nh,Hinterbichler:2011tt}. From a particle physics point of view, general relativity describes a massless spin two particle, the graviton, in interaction with matter and itself. The consistency of these interactions is guaranteed by  diffeomorphism invariance, and this allows the covariant symmetrical tensor with ten components (in 4 dimensions) to describe the two degrees of freedom of the graviton, and also leads to Einstein equations.
A generic massive deformation of the linearized Einstein equations 
leads to a massive spin two field and a scalar. In fact, the four constraints which are inherited from the Einstein equations remove four degrees of freedom leaving six degrees of freedom. In addition, the scalar couples to matter with a propagator having a sign characteristic of a pathological ghost. In the linearised theory, it 
was shown  by Fierz and Pauli \cite{Fierz:1939ix} that when 
the mass term is a particular combination of the two quadratic invariants, a new constraint arises which removes the extra degree of freedom. Since this combination is not a consequence of an additional gauge symmetry, a generic self-interaction is expected to reintroduce the extra degree of freedom, and this was confirmed by the analysis of Boulware and Deser \cite{Boulware:1973my}.
From a phenomenological point of view, the massive deformation  which seemed excluded due to the van dam Zakharov discontinuity  characteristic of the linear Fierz-Pauli theory \cite{vanDam:1970vg}
was revived by Vainstein \cite{Vainshtein:1972sx} who argued that non linearities cannot be neglected even at the solar system distance scales (for a review see \cite{Babichev:2013usa}).

A ghost-free non-linear extension of the Fierz-Pauli action was proposed in \cite{deRham:2010kj}. The absence of the ghost was first shown in the decoupling limit \cite{Creminelli:2005qk,deRham:2010ik,ArkaniHamed:2002sp} and a full Hamiltonian treatment was undertaken in \cite{Hassan:2011ea,Hassan:2011zd,Hassan:2011tf,Hassan:2011hr,Kluson,OthersCounting}.\footnote{Note, however, that these results have been debated \cite{Alberte:2010qb,Chamseddine:2011mu,Chamseddine:2013lid}.}
In its original version, the theory was formulated with two metrics \cite{Isham:1971gm} one of which, say $f$ can be fixed and the the other one, $g$ describes the massive graviton. An important aspect of the  ghost-free non-linear completion of the Fierz-Pauli mass term is that it is built from the  matrix square root of $g^{-1}f$.  This is not necessarily well defined (nor indeed real) \cite{dmz1}, but its existence can be imposed if a certain symmetry condition \cite{Chamseddine:2011mu,Volkov:2012wp,dmz2} is satisfied. In that case, the theory takes a rather simpler form in a moving basis or veilbein formulation \cite{Hinterbichler:2012cn,Nibbelink:2006sz} where  these terms become polynomial and where, for certain mass terms, the additional constraint removing the Boulware-Deser ghost is easily obtained \cite{dmz1}.  It is precisely this vielbein formulation which we study here. 

The absence of ghosts is not sufficient for a theory to be viable; for example, it was suggested 
 that the theory is acausal \cite{acausality}. Here we study  general translation-invariant time dependent fields to further explore the vielbein theory,  determine exact solutions, and finally analyse their (in)stability. (For other approaches see \cite{solutions}.)   
The framework is a simplified one, and in particular it allows one to count the degrees of freedom by solving explicitly the constraints, contrary to the case of the Hamiltonian treatment of a general space and time dependent field where the constraints cannot be explicitly solved \cite{Hassan:2011ea,Hassan:2011zd,Hassan:2011tf,Hassan:2011hr,Kluson,OthersCounting}. In the linear approximation, the homogeneous time depend solutions of the Fierz-Pauli equations are plane-wave solutions boosted to the rest frame, and the counting of the degrees of freedom is straightforwardly given by the number of polarization tensors. In the other limit of massless gravity the time dependent solutions are the Kasner \cite{Kasner:1921zz} or Bianchi I solutions to the Einstein equations in vacuum. 

The plan of the paper is as follows.
In Section \ref{FP} we review the plane wave solutions of the Fierz-Pauli theory, showing in particular that the counting of degrees of freedom is straightforward in that setting.
In Section \ref{sec:action}, we briefly review the moving basis formulation of massive gravity in order to fix notation. We concentrate on the case in which the mass term is described by a cosmological constant and a term linear in the external moving frame.
 Section \ref{sec:tt} is devoted to the  analysis of general translation invariant and time dependent solutions. An ADM-like decomposition is used, and we show that the  constraints lead to {\it i)} the vanishing of the shift vectors, {\it ii)} the expression of the lapse function in terms of the 
 spatial moving frame, {\it iii)} the constancy of the determinant of the spatial moving frame
 and finally {\it iv)} an additional scalar constraint which leads to a phase space with a nontrivial structure.
In this section we also obtain the equations of motion and show that the initial value formulation is well posed. In Section \ref{sec:solutions} we study solutions to the constraints and equations of motion. We determine the first correction to the linearized Fierz-Pauli solution and then turn to consider exact solutions.  In three dimensions, the system reduces to the analog of a one dimensional particle in a potential.
We show that the solutions belong to different disconnected sectors depending
on the initial conditions and the signature of the spatial moving basis. When the latter  is not positive definite the 
effective potential is unbounded from below. This feature is also true  for higher dimensions.
In four dimensions we study numerically the general case of a diagonal spatial moving basis and analytically a special case corresponding to two equal eigenvalues.
We collect our conclusions in Section 6.

\section{The Fierz-Pauli time-dependent solutions}
\label{FP}

The Fierz-Pauli action for a covariant tensor $h_{\mu \nu}$ defined on $D$-dimensional flat space-time $\eta_{\mu\nu}$ is given by
\be
S=-M^2 \int d^D x \Big\{ \left[ (\partial_\mu h_{\nu \rho})^2 - (\partial_\mu h)^2 + 2 (\partial_\mu h)(\partial^\nu h_{\nu}^\mu) - 2 (\partial_\mu h_{\nu \rho})(\partial^\nu h^{\mu \rho}) \right] + m^2 (h_{\mu \nu}h^{\mu \nu} - h^2) \Big\}
\label{SFP}
\ee
where $\mu=0\ldots D-1$,  $M$ is the Planck mass, and the metric signature is mostly plus.
The terms in square brackets originate from the expansion of the Einstein-Hilbert Lagrangian density to second order in $h_{\mu \nu}$ around flat Minkowski space (indices are raised and lowered with $\eta_{\mu\nu}$, so that $h=h_{\mu \nu}\eta^{\mu \nu}$). The specific combination multiplying the mass parameter $m$ is chosen so as to give -- in a consistent and ghost-free way -- a mass $m$ to the graviton $h_{\mu \nu}$.
Indeed, on varying with respect to $h_{\mu \nu}$ and using the Bianchi identities, the equations of motion take the
the Klein-Gordon form
\be 
(\Box-m^2)h_{\mu\nu}=0,
\label{FP1}
\ee
which must be solved together with the $D+1$ constraints
\be 
\partial^\mu h_{\mu\nu}=0,\quad h=0.
\label{FP2}
\ee
Thus $h_{\mu\nu}$ is a massive spin 2 field with $\frac{D(D+1)}{2}-(D+1)=\frac{D(D-1)}{2}-1$ degrees of freedom.

Here we consider solutions which depend only on time.  In that case (\ref{FP1}) and (\ref{FP2}) reduce to
\be 
\ddot{h}_{\mu\nu}+m^2h_{\mu\nu}=0,\qquad \dot{h}_{0\mu}=0,\qquad h=0
\ee
where the dot denotes a time derivative, and hence it follows that $h_{\mu 0}=0$.
The solution of these equations is readily found as
\be 
h_{\mu 0}=0,\quad  h_{ij}=\epsilon_{ij}\cos(mt+\phi),\quad \epsilon^{i}{}_i=0
\ee
with $\epsilon_{ij}$ and $\phi$ constants, and $i,j=1,\ldots,D-1$.
The degrees of freedom are thus encoded in the $(D-1)$-dimensional symmetric and traceless rank 2 tensor $\epsilon_{ij}$. Hence we find the correct $D(D-1)/2-1$ degrees of freedom of a massive spin 2 particle.

A general plane wave solution is  obtained by performing a boost on this solution.
This simple exercise shows that the degrees of freedom are simply exhibited from translation invariant and time-dependent solutions.
In the following we generalise this solution to the nonlinear interacting case, and hence we are also able to consider anisotropic Kasner-type cosmological solutions to non-linear massive gravity.

\section{Action, equations of motion, and constraints in non-linear massive gravity}
\label{sec:action}

Consider a space-time of dimension $D$ with a dynamical metric $g_{\mu\nu}$ and a non-dynamical one $f_{\mu\nu}$. 
Let $\theta^{A}$ and $f^{A}$ to be the families of 1-forms defined at every point in space-time by
\bea
&\eta_{AB}\theta^{A}{}_{\mu}\theta^{B}{}_{\nu}=g_{\mu\nu}\ ,
\label{gdef}
\\
&\eta_{AB}f^{A}{}_{\mu}f^{B}{}_{\nu}=f_{\mu\nu}\ ,
\eea
where the Lorentz indices $A,B=0,\ldots D-1$ are raised and lowered with the Minkowski metric $\eta_{AB}$.
The dual vectors $e_{A}$ to the 1-forms $\theta^{A}$ verify 
\be
\theta^{A}(e_{B})=\theta^{A}{}_{\mu}e_{B}{}^{\mu}=\delta^{A}{}_{B},
\label{inv}
\ee
and we define the $(D-n)$-forms by \cite{DuboisViolette:1986ws} 
\be
\theta^*_{A_1\dots A_n}\equiv{1\over (D-n)!}\epsilon_{A_1\dots A_d}
\theta^{A_{n+1}}\wedge\dots\wedge \theta^{A_{d}}\ .
\ee

If the symmetry property
\be
e^{C}{}^{\mu} f^{B}{}_{\mu} = e^{B}{}^{\mu} f^{C}{}_{\mu}
\label{sym}
\ee
holds,
then it has been shown that the action for the ghost-free, metric, formulation of massive gravity developed in \cite{deRham:2010kj,deRham:2010ik,Hassan:2011hr} can be elegantly written as \cite{Hinterbichler:2012cn} (see also \cite{dmz1} and references therein)
\be
S={1\over 2}\int \Omega^{AB}\wedge\theta^*_{AB}+\sum_{n=0}^{D-1}\beta_n\int f^{A_1}\wedge\dots\wedge f^{A_n}\wedge\theta^*_{A_1\dots A_n}\ ,
\label{action}
\ee
where the $\beta_n$ are arbitrary parameters, and
\be
\Omega^{AB}\equiv d\omega^{AB}+\omega^{A}{}_C\wedge\omega^{CB}\
\ee
is the curvature 2-form associated to the spin-connection $\omega^{AB}$ (which itself results from the torsion-free condition $\cD\theta^A\equiv d\theta^A+\omega^{A}{}_B\wedge\theta^B=0$ and the antisymmetry in the indices $A$ and $B$).
The curvature 2-form satisfies the Bianchi identity
\be
\cD\Omega^{AB}\equiv d\Omega^{AB}+\omega^{A}{}_{C}\wedge\Omega^{CB}+\omega^{B}{}_{C}\wedge\Omega^{AC}=0\ .
\ee
The importance of (\ref{sym}) is that it is sufficient \cite{Chamseddine:2011mu,Volkov:2012wp,dmz2} to allow the existence of a real square-root of the matrix $g^{-1}f$. It is precisely this  which appears in the potential in the metric formulation of massive gravity, and which then allows the action to be rewritten in terms of forms as (\ref{action}).  However, it should be stressed that we now take (\ref{action}) as our starting point (that is, we do {\it not} impose by hand the condition (\ref{sym}) on the vielbeins). As shown in \cite{dmz1}, for some $\beta_n$ this condition is imposed dynamically, though this is not always necessarily the case. In fact, the class of theories of massive gravity described by (\ref{action}) is larger than that of the metric formulation, and it potentially has a larger space of solutions.

The action (\ref{action}) breaks both diffeomorphism and local Lorentz invariance if the non-dynamical vielbein $f^A$ is fixed. In the following we choose $f^{A}\equiv dx^{A}$, which is always possible when $f_{\mu \nu}=\eta_{\mu \nu}$.  Thus $f^A{}_\mu = \delta^A{}_{\mu}$, and we can identify Lorentz and spacetime indices. Now we define the Einstein tensor as the $(D-1)$-form
\be
G_A\equiv -{1\over 2}\Omega^{BC}\wedge\theta^*_{ABC}\ ,
\ee
which can be decomposed in the basis of $(D-1)$-forms $\theta^*_B$ as $G_A\equiv G_{A}{}^{B}\theta^*_B$.
Then the field equations following from (\ref{action}) read
\be
G_A=t_A\ ,
\label{eom}
\ee
or equivalently $G_{AB}=t_{AB}$, with
\be
t_A\equiv \sum_{n=0}^{D-1} \beta_n f^{A_1}\wedge\dots\wedge f^{A_n}\wedge\theta^*_{AA_1\dots A_n}\equiv
t_{A}{}^{B}\theta^*_B\ .
\ee
Notice that as a result of diffeomorphism invariance of the Einstein-Hilbert term, Bianchi identity
\be
\cD G_A=0=\cD t_A\ 
\label{bb}
\ee
 must hold, whilst Lorentz invariance imposes 
\be
G_{[AB]}=0=t_{[AB]} \ .
\ee

In the following we set $\beta_n=0$ for $n \geq 2$. Thus we consider the simplest case ($\beta_0, \beta_1 \neq 0)$
leading to consistent equations of motion for a massive spin 2. 
As shown in \cite{dmz1}, the equations of motion then {\it imply} that the symmetry condition (\ref{sym}) holds,
whilst the Bianchi identity (\ref{bb}) leads to the $D$ constraints
\be
\omega^A{}_{BA}=0.
\label{conB}
\ee
Here we have decomposed the spin-connexion 1-form as $\omega^A{}_B=\omega^A{}_{BC}\theta^C$.
This vector constraint can be written in a more convenient form by using the structure functions
$C_{AB}{}^C$ defined by
\be 
[e_A,e_B]\equiv C_{AB}{}^Ce_C
\ee 
and which are given by
\be 
C_{AB}{}^C=e_{[A}{}^\mu\partial_\mu e_{B]}{}^\nu\theta^C{}_\nu \, .
\label{Cdef}
\ee
In terms of these, the spin connexion components are given by
\be 
\omega_{ABC}={1\over 2}(C_{ABC}+C_{ACB}+C_{CBA}),
\label{cmptw}
\ee 
where the indices are lowered with the Minkowski metric.
It then follows that the constraint (\ref{conB}) reads
\be 
\partial_\mu\left({e_B{}^\mu\over {\det{e}} }\right)=0.
\label{vconstraint}
\ee

Using this vectorial constraint, the equations of motion (\ref{eom}) simplify and  are given by
\be 
e_C{}^\mu\partial _\mu\omega^{C}{}_{AB}-\omega^{C}{}_{AD}\omega^{D}{}_{BC}
-{1\over 2}\eta_{AB}\omega^{C}{}_{DE}\omega^{DE}{}_{C}={1\over 2}\left[(\beta_0+\beta_1e_C^C)\eta_{AB}-\beta_1 e_{AB}\right]
\label{eom2}
\ee
where 
\be
e_{AB} \equiv e_A{}^\mu\eta_{\mu B} = e_{BA},
\ee
and the last equality results from the symmetry condition (\ref{sym}).
The coefficient $\beta_0$ is chosen such that the Minkowski metric (or $e_{AB}=\eta_{AB}$) is a solution,
\be 
\beta_0=-(D-1)\beta_1,
\ee
whilst $\beta_1$ is related to the mass of the graviton by $\beta_1=-2m^2$. 
On taking the trace of the equation of motion (\ref{eom2}) one finds \cite{dmz1} a further independent scalar constraint
\be 
{(2-D)}\omega^{C}{}_{DE}\omega^{DE}{}_{C}=-2m^2{(D-1)}\left(e_C^C-D\right) \, .
\label{scalcon}
\ee
This removes the ghost generically present in massive deformations of gravity.
Finally, on using the scalar constraint back in the equation of motion (\ref{eom2}), these simplify further and read  
\be 
e_C{}^\mu\partial _\mu\omega^{C}{}_{AB}-\omega^{C}{}_{AD}\omega^{D}{}_{BC}
= \tau_{AB}
\label{eomf}
\ee
where
\be
\tau_{AB} = m^2\left[ e_{AB}+\eta_{AB}\left({e_C^C-2(D-1)\over D-2}\right)\right] \, .
\ee
In the linearised approximation, (\ref{eomf}) reduces to the Klein-Gordon equation.

\section{Translation-invariant and time dependent massive spin 2 fields}
\label{sec:tt}

We now look for solutions invariant under spatial translations. These will generalise the plane wave solution of the  Pauli-Fierz equations and the Bianchi I or Kasner solution of general relativity.

A convenient ADM-type splitting of the moving frame components is to write the symmetric matrix $e_{AB}$ as
\be 
e_{00}=-N\, ,\qquad e_{0i}=-Nn_i \, ,\qquad e_{ij}=\pi_{ij}-Nn_in_j \, ,
\ee
where the $D(D+1)/2$ variables all depend on $t=x^0$. We also define $\theta^{AD} = \theta^{A}{}_{\mu} \eta^{\mu D}$, so that from (\ref{inv}), $e_{AB} \theta^{BC} = \delta^{C}_{\; A}$. Hence
\be 
\theta^{00}=-\left({1\over N}-(\pi^{-1})^{ij}n_in_j \right)\, ,\qquad   \theta^{0i}=-(\pi^{-1})^{ij}n_j \,, \qquad \theta^{ij}=(\pi^{-1})^{ij},
\ee
where $(\pi^{-1})^{ij}$ is the inverse of $\pi_{ij}$.

\subsection{The constraints} 
Since $\det e=N\det\pi$, it follows that time component of the  vector constraint (\ref{vconstraint}) implies 
\be 
\partial_t\det\pi=0 
\label{detpiconst}
\ee
whilst the spatial components then imply that
\be 
\partial_t n_i=0.
\ee
Thus both $\det\pi$ and the `shift' vector $n_i$ are constant in time.  Notice that (\ref{detpiconst}) can also be rewritten as
\be
\tr(\pi^{-1} \partial_t {\pi})=0 \,.
\label{Giovanni}
\ee
In order to determine the scalar constraint as well as the equations of motion, we need components of the spin connexion. These are straightforward to calculate: on defining
\be 
\alpha_{i}{}^j=(\partial_t \pi_{ik})(\pi^{-1})^{kj}
\label{alphadef}
\ee
we find
\bea
\omega_{ijk}&=&{N\over 2}(n_i\alpha_{(jk)}-n_j\alpha_{(ik)}-n_k\alpha_{[ij]})
\\
\omega_{ij0}&=&{N\over 2}\left(n_{[i}\alpha_{j]}{}^\ell n_\ell-\alpha_{[ij]}\right),\\
\omega_{0ij}&=&{N\over 2}\left(\alpha_{(ij)}-n_{[i}\alpha_{j]}{}^\ell n_\ell\right) \\
 \omega_{0i0}&=&N
\alpha_{i}{}^{\ell}n_\ell \, ,
\eea
where $\alpha_{(ij)} =\alpha_{ij} + \alpha_{ji} $ and $\alpha_{[ij]} =\alpha_{ij} - \alpha_{ji}$, and from now on spatial indices are raised and lowered with $\delta_{ij}$.
It then follows from (\ref{scalcon}) that, in matrix notation, the scalar constraint reads
\bea
\frac{N^2}{2} Q
=2m^2 \left({D-1\over D-2}\right) \left[N(1-n^2)+\tr(\pi) -D\right] \, ,
\label{consts}
\eea
where $n^2 = n_i n^i = n^T n$ and
\be
Q \equiv  {1\over2} (1-n^2)\tr \left((\alpha+\alpha^T)^2\right)+n^T\left(\alpha\alpha^T-\alpha^T\alpha\right)n
+n^2 \left(n^T\alpha^T\alpha n\right) -(n^T\alpha n)^2 \, .
\label{Qdef}
\ee

On the other hand, the equations of motion (\ref{eomf}) become
\be 
N\partial_t\tilde\omega_{AB}-\omega^{C}{}_{AD}\omega^{D}{}_{BC}=\tau_{AB},
\label{eofmf}
\ee
where
\be 
\tilde \omega_{AB}=\omega^0{}_{AB}+n_i\omega^i{}_{AB},
\ee
so that
\bea 
\tilde\omega_{00}&=& n^j \omega_{j00} = -Nn^i\alpha_{ij}n^j
\\
\tilde\omega_{0i}&=&n^j \omega_{j0i} = -{N\over 2} n^j\left(\alpha_{(ji)} - n_{[j}\alpha_{i]\ell} n^\ell \right) = \tilde\omega_{i0}
\\
\tilde\omega_{ij}&=&-\omega_{0ij}+ n^k \omega_{kij} = {N\over 2}\left[ (n^2-1)\alpha_{(ij)}-n^k\alpha_{k(i}n_{j)}\right] \, .
\eea
We will study these equations of motion in section \ref{sec:eofm}.  Notice, however, that $D$ more constraints can be obtained from the $\tilde{\omega}_{AB}$, since these  satisfy
\be 
\tilde\omega_{00}-n^i\tilde\omega_{0i}=0,\qquad \tilde\omega_{0i}-n^j\tilde\omega_{ij}=0.
\ee
Thus, on taking the corresponding combinations of the equations of motion, one finds
\bea 
n^i\omega^{C}{}_{0D}\omega^{D}{}_{iC}-\omega^{C}{}_{0D}\omega^{D}{}_{0C}&=&\tau_{00}-n^i\tau_{0i},
\label{c1}
\\
n^j\omega^{C}{}_{iD}\omega^{D}{}_{jC}-\omega^{C}{}_{0D}\omega^{D}{}_{iC}&=&\tau_{0i}-n^j\tau_{ij}.
\label{c2}
\eea
A rather involved calculation enables the first constraint (\ref{c1}) to be rewritten as
\be
\frac{N^2}{2}Q = m^2\left[\frac{(D-1)(N(1-n^2)-2)+\tr(\pi)}{ D-2}\right]
\label{consts2}
\ee
where $Q$ is given in (\ref{Qdef}), whilst the second one takes the form
\be
\frac{N^2}{2}Q n_i=m^2\left\{ \left[\frac{(D-1)(N(1-n^2)-2)+\tr(\pi)}{ D-2}\right] n_i-(\pi n)_i \right\}
\label{constv2} \, .
\ee
On comparing (\ref{consts2}) and (\ref{constv2}), it follows that $(\pi n)_i=0$ and hence $n_i=0$ since $\pi_{ij}$ must be an invertible matrix.

The system of constraints for time-dependent solutions of massive gravity then simplifies considerably. In particular $Q$ takes the simple expression
\be
Q = {1\over2}\tr \left((\alpha+\alpha^T)^2\right)
\ee
so that the ratio of (\ref{consts2}) and (\ref{consts}) yields
\be 
N-1=-\left({2D-3\over D-1}\right) \tr(\pi-\mathbb{1}),
\label{linkNpi}
\ee
showing that $N$ and $\tr(\pi)$ are directly related.
Finally (\ref{consts2}) yields the two identities
\bea
2m^2\tr(\pi-\mathbb{1})=-{N^2\over 4}\tr \left((\alpha+\alpha^T)^2\right),
\label{use1}
\\
2m^2(N-1)=\frac{N^2}{4}\left({2D-3\over D-1}\right) \tr \left((\alpha+\alpha^T)^2\right).
\eea
Thus once $\tr(\pi)$, say, is known, then both $N$ and $\tr\left((\alpha+\alpha^T)^2\right)$ can be determined.
Finally, note that imposing that $N$ should be real leads to the inequalities
\be 
{2D-3\over 2(D-1)}\tr \left((\alpha+\alpha^T)^2\right)\leq m^2\, , \qquad
\tr(\pi - \mathbb{1})< 0.
\label{trpi}
\ee

To summarise, the outcome of the constraints analysis is the following. The `shift' $n_i$ vanishes whilst the matrices $\pi$ and $\partial_t{\pi}$ are subject to
\bea
\tr[(\partial_t{\pi})\pi^{-1}]&=&0
\label{ps1} \\
2m^2\tr(\pi-\mathbb{1})&=&-{1\over 4}\left(1-{2D-3\over D-1}\tr(\pi-\mathbb{1})\right)^2\tr\left\{ \left[ (\partial_t{\pi})\pi^{-1}+\pi^{-1}(\partial_t{\pi}) \right]^2\right\} \, .
\label{ps2}
\eea
Finally the shift $N$ is determined from $\pi$ from (\ref{linkNpi}).

Since $n_i$ vanishes for all time, and since $(N,\partial_t{N})$ are determined from $(\pi,\partial_t{\pi})$ (from equation (\ref{linkNpi}) and its derivative), it follows that the phase space of this system is given by the set of matrices $\pi$ and $\partial_t{\pi}$.  This phase space has a very non-trivial, non-linear, structure due to the two constraints (\ref{ps1}) and (\ref{ps2}).  Its dimension, though, is $D(D-1)-2$ as it should be in order to describe a spin 2 massive particle. One of the degrees of freedom is just the constant $\det{\pi}$.

It is interesting to see how this non-trivial structure reduces to that of the Fierz-Pauli system considered in section \ref{FP} by linearising the constraints around the identity $\pi= \mathbb{1}
+h$.  These then reduce to
\be 
n_i=0,\qquad\tr(\partial_t{h})=0,\qquad\tr({h})=0,\qquad N=1.
\label{linPFF}
\ee
Thus $N$ is independent of $h$, whilst the second constraint is a consequence of the third.   Notice that in general, however, (\ref{ps1}) is not a consequence of (\ref{ps2}).

\subsection{The equations of motion}
\label{sec:eofm}

We now turn to the equations of motion.  Since $n^i=0$, first notice that the dynamical metric $g_{\mu \nu}$ defined in (\ref{gdef}) takes the form
\be
ds^2 = g_{\mu \nu}dx^\mu dx^\nu = - \frac{1}{N^2} dt^2 + (\pi^{-2})_{ij} dx^i dx^j \, .
\ee
It is therefore convenient to define a new time coordinate $T$ by 
\be
dT=\frac{1}{N}dt,
\label{Tt}
\ee
and the matrix $B_{ij}$ by
\be
B \equiv \pi^2.
\label{Bdef}
\ee 
Note that since $\tr(\pi - \mathbb{1})< 0$ (see (\ref{trpi})) it follows from (\ref{linkNpi}) that $N>1$, and hence $t(T)$ is unambiguously determined from (\ref{Tt}).

The only remaining non-zero component of the equations of motion (\ref{eofmf}) is the $ij^{\rm{th}}$ one, which reads
\be 
-{1\over 2}\partial_T(N\alpha_{(ij)})-{N^2\over 4}\left(\alpha_{(ik)}\alpha^{[k}{}_{j]}+
\alpha_{(jk)}\alpha^{[k}{}_{i]}\right)=m^2\left( \pi_{ij}-{\delta_{ij}\over D-1}\tr\pi \right) \, ,
\label{eofmff}
\ee
where we have simplified the right hand side using (\ref{linkNpi}).  
This takes a more illuminating form when expressed in terms of the symmetric and anti-symmetric matrices $S$ and $A$ defined respectively by (see (\ref{alphadef}) for the definition of $\alpha$ in terms of $\pi$), 
\bea
S&\equiv& {N\over 2}(\alpha+\alpha^T) = {1\over 2}\{\dot{\pi},\pi^{-1}\}
\\
A&\equiv& {N\over 2}(\alpha-\alpha^T) = {1\over 2}\, [\dot{\pi},\pi^{-1}] \, ,
\eea
where here and in the following $\cdot = \partial_T$. 
Then (\ref{eofmff}) becomes
\be 
\dot{S}+[S,A]=-m^2\left(\pi-{\mathbb{1}\over D-1}\tr(\pi)\right) \, ,
\label{Fede}
\ee
whilst the constraints (\ref{Giovanni}) and (\ref{use1}) become 
\be
\tr{S} = 0 \, , \qquad \tr\left(S^2+2m^2(\pi-\mathbb{1})\right)=0 .
\label{FedeZ}
\ee
These are of course compatible with (\ref{Fede}), since taking the trace gives
$\partial_T\tr S=0$, while the anticommutator  with $S$ gives
$\partial_T(\tr S^2+2m^2\tr\pi)=0$.

The equation of motion also simplifies when written in terms of the matrix $B=\pi^2$. Since 
 \bea
\partial_T(B^{-1}\dot{B})&=& \partial_T(2\pi^{-1}S\pi)
=
2\pi^{-1}\Big(\dot{S}+[S,\dot{\pi}\pi^{-1}]\Big)\pi=2\pi^{-1}\Big(\dot{S}+[S,A]\Big)\pi \, ,
\eea
an alternative and particularly useful form of (\ref{Fede}) is 
\be 
\partial_T(B^{-1}\dot{B})=-2m^2\left(\pi-{\mathbb{1}\over D-1}\tr(\pi)\right)  = \partial_T(\dot{B} B^{-1}) \,,
\label{EZ}
\ee
from which it follows directly that the antisymmetric matrix
\be 
\beta\equiv {1\over2} [\dot{B},B^{-1}]
\label{betadef}
\ee
is {\it conserved}, $\dot{\beta}=0$.  Thus our system contains $(D-1)(D-2)/2$ conserved quantities, which correspond to the expected conservation of angular momentum since the system is invariant under spatial rotations.  These will be used extensively below when we search for exact solutions. Finally, in terms of $B$, the constraints (\ref{FedeZ}) read
\be 
\tr(B^{-1}\dot{B})=0,\qquad \tr\Big((B^{-1}\dot{B})^2+8m^2(\pi-\mathbb{1})\Big)=0.
\label{conSP}
\ee

To summarise, phase space $(\pi,\dot{\pi})$ contains the $D(D-1)-2$ degrees of freedom necessary to describe a massive spin 2 field. Of those, $(D-1)(D-2)/2 + 1$ are constants of motion coming from the conservation of $\beta$ and the conservation of $\det(\pi)$.  For example, in $D=3$ dimensions, there are $4$ degrees of freedom, of which 2 are constants of motion: we will study this situation in section \ref{sec:D3}.

\subsection{The initial value formulation}
\label{sec:dof}
Before turning to exact solutions (section \ref{sec:solutions}), we first address the initial value problem.   
Notice that $\ddot{\pi}$ appears in the equations of motion through the combination
$\ddot{\pi}\pi^{-1}+\pi^{-1}\ddot{\pi}$ (which is in $\dot{S}$). It is therefore not manifest that the equations of motion give $\ddot{\pi}$ in terms of lower derivatives --- or in other words, it is not clear that given $\pi$ and $\dot{\pi}$ satisfying the constraints at a given time, the equations determine in a unique way $\pi$ and $\dot{\pi}$ at a later time.  We now show that indeed this is the case.

To do so, it is simplest to write the real symmetric matrix $\pi(T)$ in the form
 \be
\pi(T)=\sum_i\epsilon_i e^{f_i(T)}|v_i(T)\rangle\langle v_i(T)|,
\ee
where the $|v_i(T)\rangle$ are its orthonormal eigenvectors with eigenvalues $\epsilon_ie^{f_i}$ where $\epsilon_i=\pm1$ ($i=1,\ldots,D-1$). Hence
\bea 
B(T)&=&\pi(T)^2 = \sum_i e^{2f_i(T)}|v_i(T)\rangle\langle v_i(T)|
\nonumber
\eea
so that
\be
{1\over 2}\{B^{-1},\dot{B}\}=\sum_{i,j}\left[ 2\dot{f_i}\delta_{ij}+\sinh{2(f_i-f_j)}\langle v_j(T)|\dot{v_i}(T)\rangle  \right]    \ |v_j(T)\rangle\langle v_i(T)|,
\label{FDL}
\ee
and the conserved antisymmetric matrix $\beta$ defined in (\ref{Bdef}) takes the form 
\bea 
\beta&\equiv& \frac{1}{2} [\dot{B},B^{-1}] = 
2\sum_{i,j}\sinh^2{(f_i-f_j)}\langle{v_i}(T)|\dot{v_j}(T)\rangle|v_j(T)\rangle\langle v_i(T)|.
\label{FSP}
\eea
The matrix elements of $\beta$ in a time independent basis give $(D-1)(D-2)/2$ constants of motion. If we choose that basis to be the eigenvectors $|v_a(0)\rangle$at the initial time $T=0$, these are given by
\bea 
\beta_{ab}&=&2\sum_{i,j}\sinh^2{(f_i-f_j)}\langle{v_i}(T)|\dot{v_j}(T)\rangle\langle v_a(0)|v_j(T)\rangle\langle v_i(T)|v_b(0)\rangle\nonumber\\&=&2\sinh^2{(f_a-f_b)(0)}\langle v_b(0)|\dot{v_a}(0)\rangle. 
\eea
From (\ref{FSP}) it also follows that
\be 
\langle{v_i}(T)|\dot{v_j}(T)\rangle={\langle v_j(T)|\beta|v_i(T)\rangle\over 2\sinh^2{(f_i-f_j)}} \,  \qquad (i \neq j).
\label{bet}
\ee

We can now write down the constraints. From (\ref{conSP}), and on using (\ref{FDL}) as well as the conservation of $\beta$ in (\ref{bet}), the constraints become give
\be  
\sum_i \dot{f_i}=0, 
\label{conSP}
\ee
and
\be 
4\sum_i\dot{f_i}^2+\sum_{i,j \neq i}\left({\langle v_j(T)|\beta|v_i(T)\rangle\over \sinh{(f_i-f_j)}}\right)^2
+8m^2\sum_i\left(\epsilon_ie^{f_i}-1\right)=0.
\label{conDL}
\ee
The equations of motion (\ref{EZ}) reduce to 
\bea
&&2\ddot{f_i}-\frac{1}{2}\sum_{j \neq i}\sinh{2(f_i-f_j)}\left( 
{\langle v_j(T)|\beta|v_i(T)\rangle\over \sinh^2{(f_i-f_j)}}
\right)^2=-2m^2\Big(\epsilon_ie^{f_i}-{1\over D-1}\sum_j\epsilon_je^{f_j}\Big),
\label{eqm1}
\eea
and  $\forall \; i\neq j$,
\bea
&&\partial_T\big(\sinh{2(f_i-f_j)}\langle v_j(T)|\dot{v_i}(T)\rangle\big) +2(\dot{f_i}-\dot{f_j})\langle{v_j}(T)|\dot{v_i}(T)\rangle
\nonumber
\\
&& \qquad-\sum_k\left[ \sinh{2(f_i-f_k)}+\sinh{2(f_j-f_k)}\right]\langle v_k(T)|\dot{v_i}(T)\rangle\langle v_k(T)|\dot{v_j}(T)\rangle
=0 .\ 
\label{eqm2}
\eea
One can verify that the time derivative of (\ref{conDL}) is identically satisfied as a concequence of (\ref{eqm1}) and (\ref{eqm2}).
Furthermore (\ref{eqm2}) equation is redundant since it is identically satisfied once $\langle v_j|\dot{v_i}\rangle$ is replaced by its expression in (\ref{bet}).

We are now in position to analyse the initial value formulation.
Suppose that at $T=0$, we are given $\pi(0)$ and $\dot{\pi}(0)$, and hence equivalently $f_i(0)$, $\dot{f_i}(0)$, $|v_i(0)\rangle$ and 
$|\dot{v_i}(0)\rangle$ that satisfy the two constraints (\ref{conSP}) and (\ref{conDL}). From these determine the conserved quantities $\beta$, and  obtain $f_i(\delta T)$, $|v_i(\delta T)\rangle$. From (\ref{bet}) we determine $|\dot{v_i}(\delta T)\rangle$ and from (\ref{eqm1}) we get $\dot{f_i}(\delta T)$. So (\ref{bet}) and (\ref{eqm1}) are sufficient to solve completely the system once a correct set of initial values satisfying the constraints is given.

Note from (\ref{conSP}) that the ``potential" is unbounded from below if at least one of the eigenvalues is negative, signaling an instability. Notice also that when $\beta=0$, the eigenvectors $|v_i(T)\rangle$ are constant, and $B(T)$ is diagonal for all $T$.  We will discuss examples of these two situations below.

\section{Solutions}
\label{sec:solutions}

In this section we search for solutions to the equations (\ref{EZ})
\bea
\frac{1}{2}\partial_T\left(\{ B^{-1},\dot{B}\}\right)&=& -2m^2\left(\pi-{\mathbb{1}\over D-1}\tr\pi\right),
\label{EZ1}
\\
{1\over2} \partial_T \left([\dot{B},B^{-1}] \right) &=& 0 = \dot{\beta} \, , 
\label{betab}
\eea
where $B=\pi^2$, subject to the constraints (\ref{conSP}), namely
\be 
\tr(B^{-1}\dot{B})=0,\qquad \tr\left((B^{-1}\dot{B})^2+8m^2(\pi-\mathbb{1})\right)=0 \, .
\label{conSP2}
\ee
First we show that there are no isotropic solutions other than flat space. Then we consider the first correction to the linearised Fierz-Pauli solution. We then solve this system exactly in $D=3$ dimensions, and finally study the case in which $B$ is a diagonal matrix in four dimensional case.

\subsection{Isotropic solution}

Here $\pi$ is proportional to the identity matrix $\pi(t)=a(t)\mathbb{1}$. The first constraint in (\ref{conSP2}) imposes that $a$ is constant, whilst the second gives $2m^2(a-1)=0$. Thus $a=1$ is the only isotropic solution.  A similar situation is well known to arise in metric massive gravity \cite{D'Amico:2011jj}.

\subsection{Perturbative solution about flat space}
\label{sec:pfs}

The first order perturbative solution was considered in section \ref{FP}.  Here we work to second order, writing
\be
\pi = \mathbb{1} +  h +  \ell
\ee
where $h$ is of order 1, and $\ell$ of order 2.

As discussed in (\ref{linPFF}), to linear order the two constraints (\ref{conSP2}) reduce to
$\tr(\dot{h})=0 =\tr({h})$, while the equation of motion is $\ddot{h} + m^2 h=0$. Their solution is 
\be
h(T)=\epsilon\cos(mT+\phi)
\label{hPF}
\ee
with $\epsilon$ a traceless tensor. Furthermore to this order the lapse $N=1$ (see (\ref{linPFF})), so we can replace $t$ by $T$. 
To quadratic order
the two constraints and equation of motion read respectively
\be
 \tr(\dot{h}h) = \tr(\dot{\ell}) \, , \qquad \tr(\dot{h}^2)=-2m^2 \tr(\ell)\, ,  \qquad h\ddot{h} + \dot{h}^2 - \ddot{\ell} = m^2\left(\ell - \frac{\mathbb{1}}{D-1}\tr(\ell) \right) \, .
 \label{set}
 \ee
 On using (\ref{hPF}) the set of equations (\ref{set}) are solved by
\be
\ell(T) = -\frac{\mathbb{1}}{4(D-1)} \tr(\epsilon^2) + \frac{1}{3} \left(\epsilon^2-{\mathbb{1}\over 4(D-1)}\tr\epsilon^2\right)
\cos(2(mT+\phi)) \, .
\ee
Again to this order, we can replace $T$ by $t$ in the above.  Notice that this second order correction remains small for all $t$, and it is straightforward to iterate this process to higher orders. In general, to an order $n$ the correction will have a contribution in $\cos(n(mT+\phi))$.

\subsection{Solutions in $D=3$ dimensions}
\label{sec:D3}

In $D=3$ dimensions the problem of solving for $\pi(T)$ and $N(T)$ is straightforward and reduces to quadratures.  From the discussion in section \ref{sec:eofm}, we know that phase space has $D(D-1)-2 = 4$ degrees of freedom, of which 2 are constants of motion, and hence the system is integrable.

Let $\pi(T)=\pi_0(T)+\pi_1(T)\sigma_1+\pi_3(T)\sigma_3$, where $\sigma_i$ are the Pauli matrices.  Then
\be
B = (\pi_0^2 + \pi_1^2 +\pi_3^2)\mathbb{1} + 2 \pi_0(\sigma_1 \pi_1 + \sigma_2 \pi_2)
\ee 
with
\be
\det(B) = \Delta^2 \equiv (\pi_0^2-\pi_1^2-\pi_3^2)^2.
\ee
The constraint $\tr(B^{-1}\dot{B})=0$ in (\ref{conSP2})  is equivalent to $\partial_T(\det(\pi))=0$ so that
\be
\Delta = \det(\pi) = {\rm constant}\,,
\ee
while conservation of $\beta={1\over2} [\dot{B},B^{-1}]$  yields directly the second conserved quantity 
\be 
(\dot{\pi_1}\pi_3-\dot{\pi_3}\pi_1)\pi_0^{2} \equiv c 
\label{consta}
\ee
where $c=\beta_{12}\Delta $. The second constraint in (\ref{conSP2}) then leads to 
\be
{1\over \Delta^2}\left[\pi_0^2(\dot{\pi_1}^2+\dot{\pi_3}^2-\dot{\pi_0}^2)-\Delta\dot{\pi_0}^2\right]+2m^2(\pi_0-1)=0.
\label{stuff}
\ee
Since the equations of motion (\ref{EZ1}) are time derivatives of the above first order equations, we do not write them out explicitly.  

\begin{figure*}
\includegraphics[width=0.45\textwidth,height=.35\textwidth]{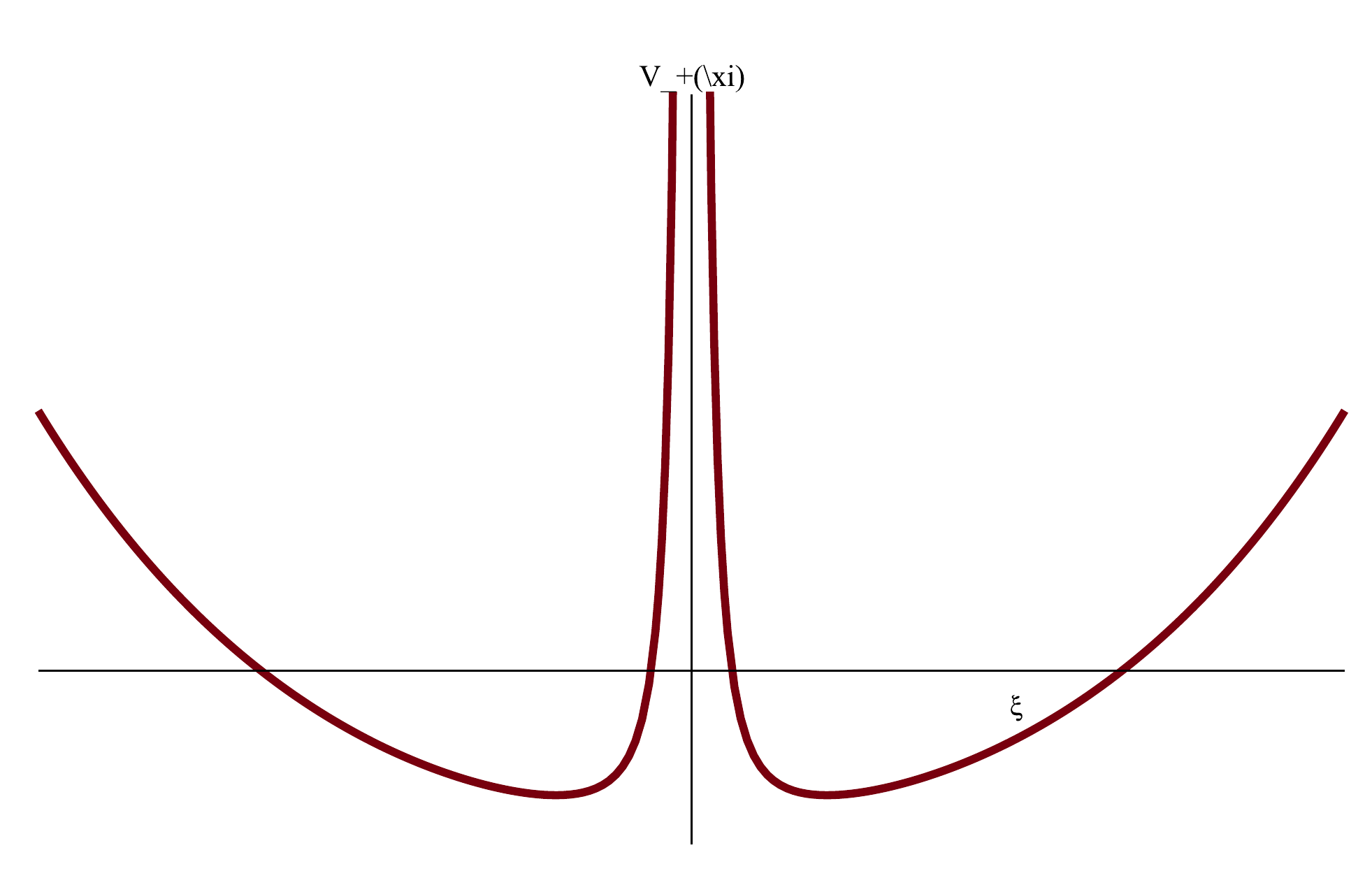}
\hspace{1cm}
\includegraphics[width=0.45\textwidth,height=.35\textwidth]{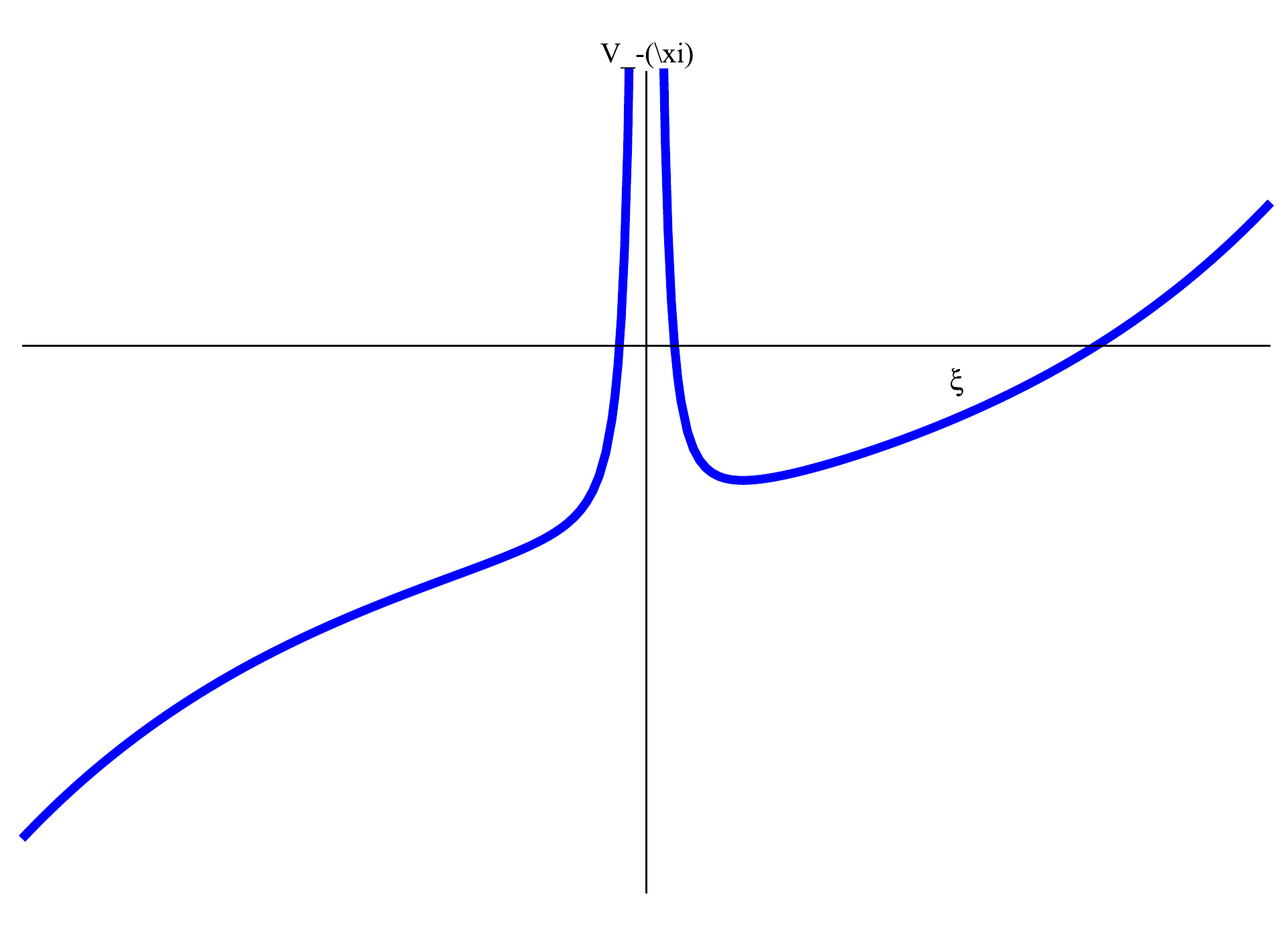}
\caption{The potentials $V_+(\xi)$ and $V_-(\xi)$ [left and right panels respectively], for $\ell \neq 0$}
\label{fig:pot}
\end{figure*}

The solutions for the $\pi_i$ depend on the sign of the constant $\Delta$.
If $\Delta >0$, a convenient change of variables is 
\be 
\pi_0=\rho\cosh\xi,\quad \pi_1=\rho\sinh\xi\sin\theta,\quad\pi_3=\rho\sinh\xi\cos\theta, \quad {\rm with} \quad \rho^2\equiv\Delta
\ee
whilst for $\Delta <0$ we set
\be
\pi_0=\rho\sinh\xi,\quad \pi_1=\rho\cosh\xi\sin\theta,\quad\pi_3=\rho\cosh\xi\cos\theta \quad {\rm with} \quad \rho^2\equiv-\Delta.
\ee
In both cases
the conserved quantity (\ref{consta}) is given by
\be 
c=\rho^4\cosh^2\xi\sinh^2\xi\dot{\theta}\equiv\rho^4\ell ,
\label{constab}
\ee
and the quadratic constraint (\ref{stuff}) takes the form (the $\pm$ labels the sign of $\Delta$)
\be
 \dot{\xi}^2+V_{\pm}(\xi) = 0,\label{ene3}
 \ee
 with
\bea
V_+ &=& {\ell^2\over \cosh^2\xi\sinh^2\xi}+2m^2(\rho\cosh\xi -1)
\label{Vp}
\\
V_- &=& {\ell^2\over \cosh^2\xi\sinh^2\xi}+2m^2(\rho\sinh\xi -1).
\label{Vm}
\eea
The four degrees of freedom of this system are therefore $(\xi,\dot{\xi})$ and $(\theta,\dot{\theta})$. They are determined from (\ref{ene3}) and (\ref{constab}) 
in terms of the two constants of motion $\Delta$ and $c$.   The potentials $V_{\pm}(\xi)$ are plotted in figure \ref{fig:pot} for $\ell \neq 0$.   In both cases the term in $\ell^2$ acts as an angular momentum barrier at $\xi=0$ meaning that the sign of $\xi$ is conserved during the evolution. Hence there are two disconnected regions of possible solutions; which is realised will depend on the initial conditions for $\xi$.

 The potential $V_+$ is symmetric in $\xi$ it can only be negative in some range of $\xi$ provided $0 \leq \rho < 1$.  This condition is equivalent to the second one in (\ref{trpi}).
Furthermore the potential is only negative if $m^2/\ell^2$ is greater than a $\rho$-dependent critical value (calculated in the Appendix, and equivalent to the first condition in (\ref{trpi})). In that case, with appropriate initial conditions, $\xi$ has periodic motion.  If $m$ is less than the critical value there is no solution.

 The potential $V_-(\xi)$ leads to two different behaviours. If $m$ is sufficiently large, for positive $\xi$ periodic motion is again possible. For negative $\xi$ however, the potential is unbounded from below and the solution is run away with $\xi \rightarrow -\infty$ and hence $|\pi_i|\rightarrow \infty$ as $T\rightarrow \infty$.  This describes a metric $g_{\mu \nu}$ whose spatial components vanish as $T\rightarrow \infty$.  Classically this theory is not pathological,  though quantum mechanically it may suffer from instabilities.

\subsection{Diagonal solutions in $D=4$ dimensions}

We finish by looking for solutions with $\pi= {\rm diag}(\pi_1,\pi_2,\pi_3)$ for all times, in which case $\beta=0$ from (\ref{betadef}).  These generalise Bianchi-I and Kasner solutions of General Relativity.

\begin{figure*}
\includegraphics[width=0.95\textwidth,height=.45\textwidth]{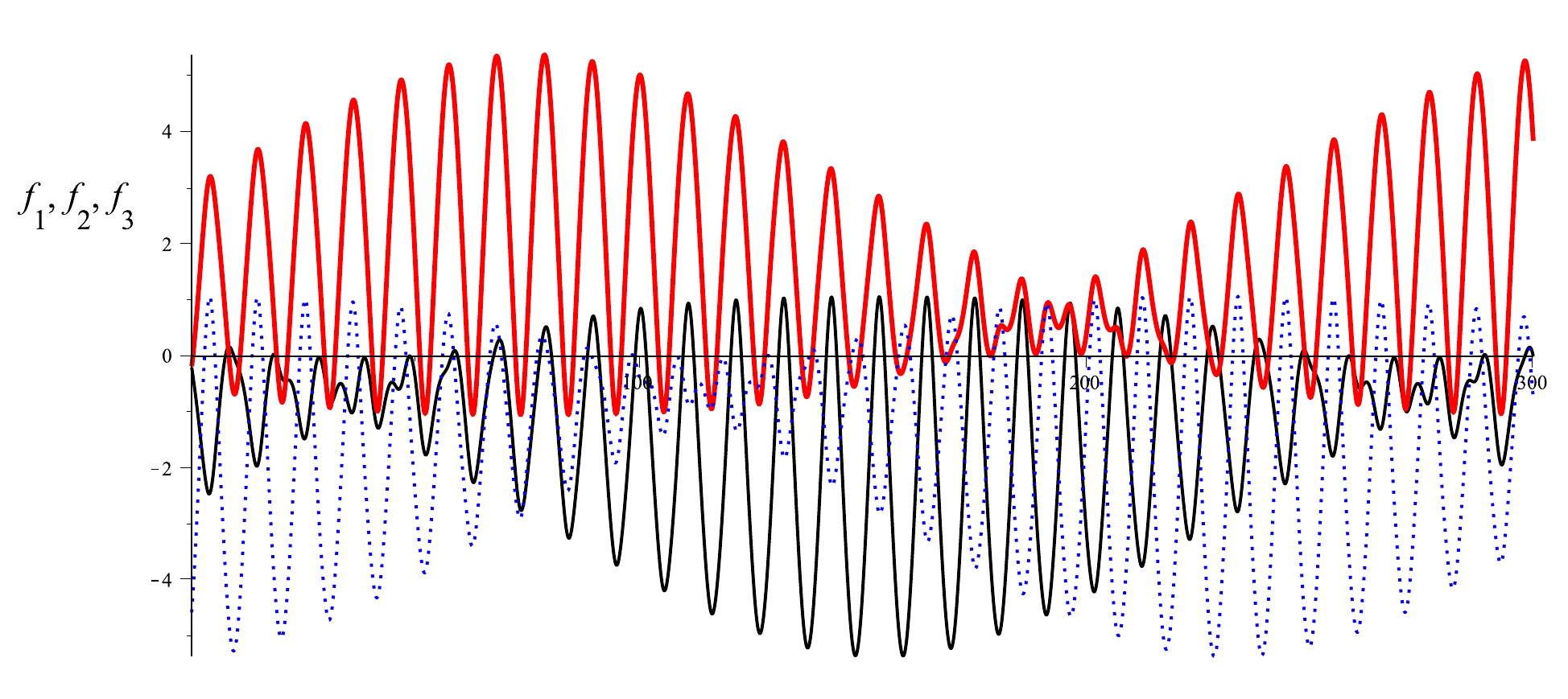}
\caption{Numerical solutions of (\ref{nearly}), showing the eigenvalues $f_1$ (black), $f_2$ (red), $f_3$ (blue, dotted) as a function of $mT$ for $\lambda = 0.01$.}
\label{fig:an}
\end{figure*}

\begin{figure*}
\includegraphics[width=0.95\textwidth,height=.45\textwidth]{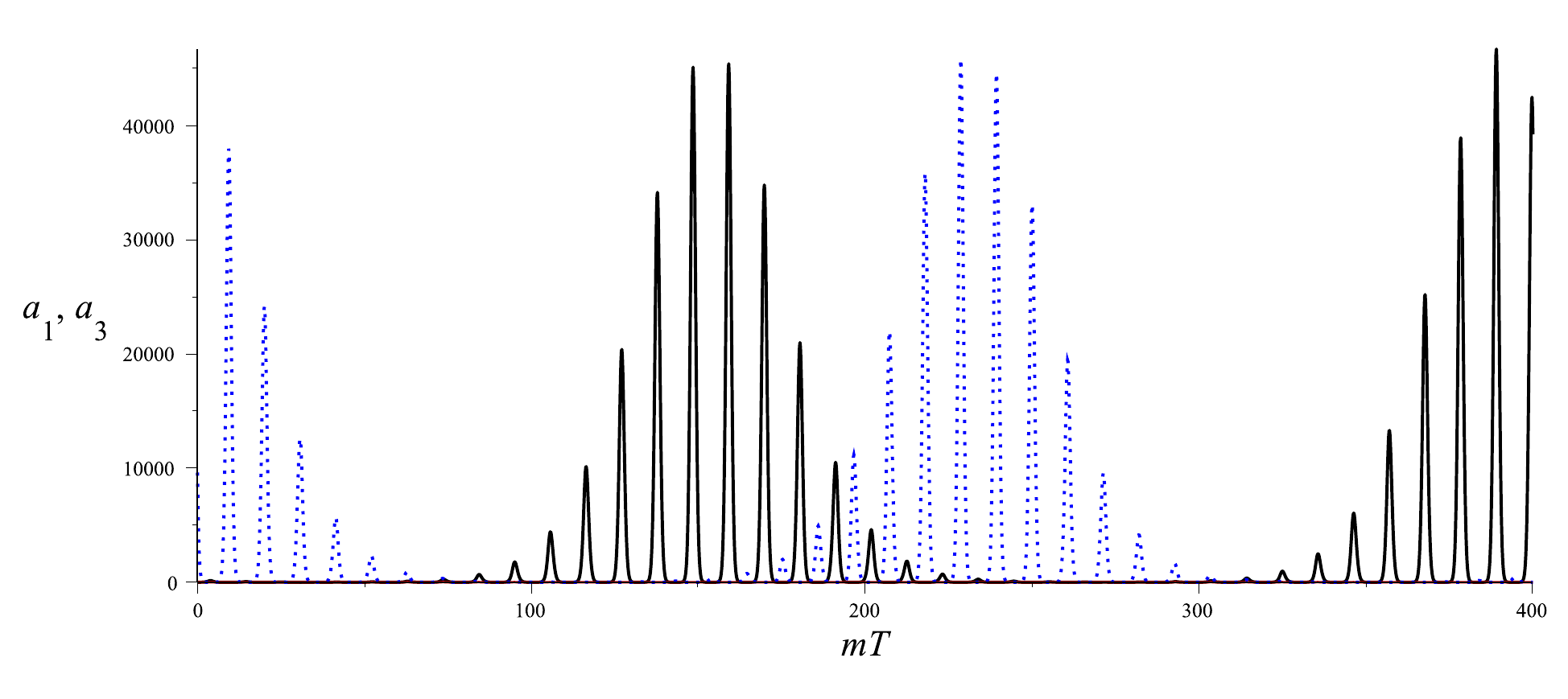}
\caption{Analogue of figure \ref{fig:an} but for the scale factors $a_i=e^{-2f_i}$. Notice from figure \ref{fig:an} that $f_{1,3}$ are mainly negative meaning that the corresponding scale factors are large.  On the other hand $f_2$ is mainly positive, so that $a_2$ is small (it is not visible on this plot).}
\label{fig:an1}
\end{figure*}

\begin{figure*}
\includegraphics[width=0.95\textwidth,height=.45\textwidth]{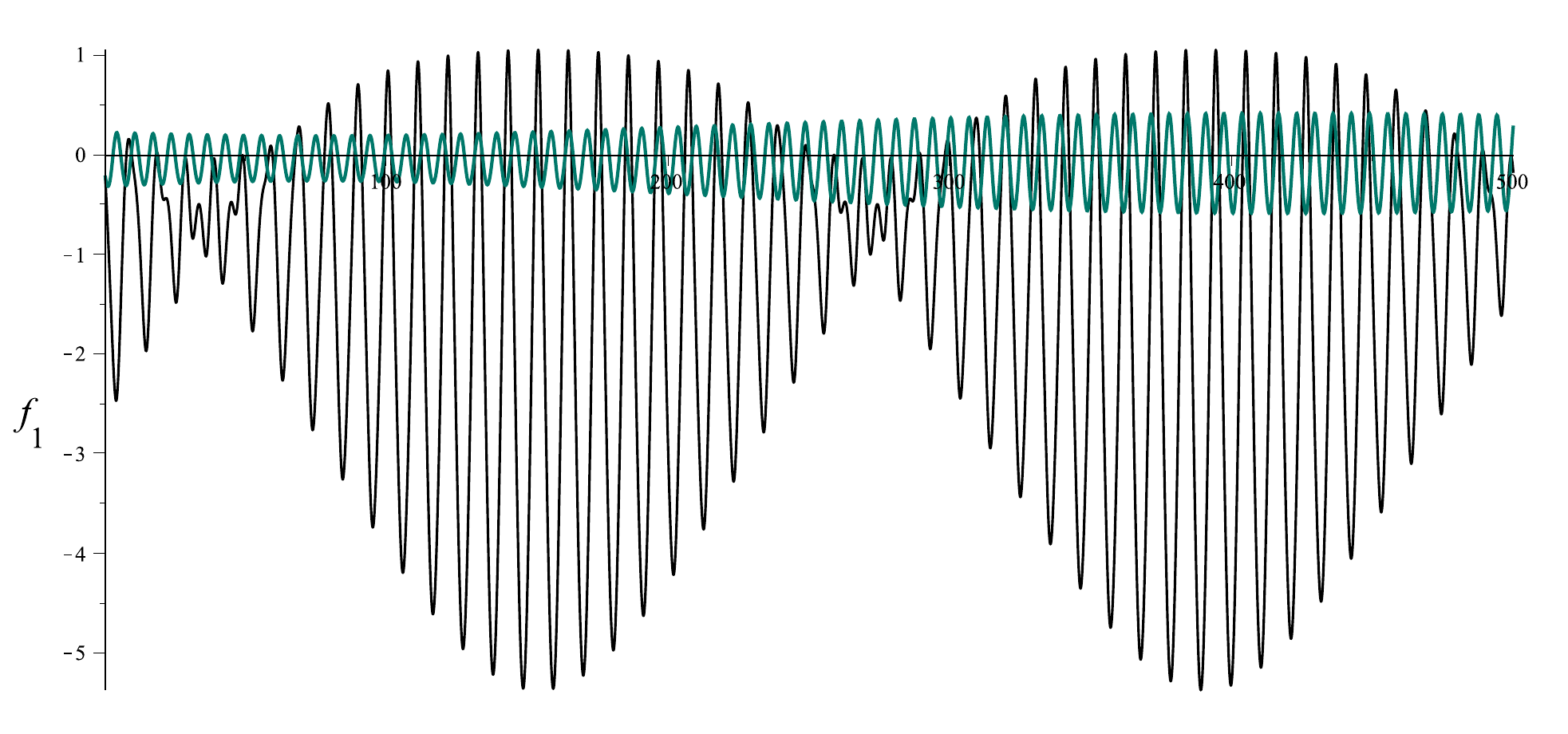}
\caption{$f_1$ as a function of $mT$ for two different values of $\lambda$.  The large amplitude (black) curves corresponds to $\lambda=0.01$. The small amplitude (green) curve has $\lambda = 0.8$.}
\label{fig:lambda}
\end{figure*}

Let $\pi_i = \epsilon_i e^{f_i}$ and $\epsilon_i = \pm 1$ is the sign of the eigenvalues of $\pi$.   Since in the following we are not interested in unbounded potentials which lead to runaway solutions, we immediately set $\epsilon_i=+1$ for all $i$.  
Then the system of equations and constraints (\ref{EZ1})-(\ref{conSP2}) reduces to
\be 
\sum_i f_i=\ln \lambda,
\ee
where $\lambda$ is the constant determinant of $\pi$, and 
\bea
3\dot{f}_+^2&+&\dot{f}_-^2+m^2\left(\lambda e^{-2f_+}+2e^{f_+}\cosh{f_-} -3\right)=0
\label{nearly}
,\\
\ddot{f}_-&=&-m^2e^{f_+}\sinh{f_-}\\
\ddot{f}_+&=&{m^2\over 3}\left(\lambda e^{-2f_+}-e^{f_+}\cosh{f_-}\right),
\eea 
where $f_{\pm}=(f_1\pm f_2)/2$.  These are the equations for a two-particle system with positions $f_+$ and $f_-$, masses $3$ and $1$, and vanishing energy in the potential
\be 
V(f_+,f_-)={m^2\over 2}\left(\lambda e^{-2f_+}+2e^{f_+}\cosh{f_-} -3\right).
\ee
The potential is minimum when $f_-=0$, and can only take negative values for $0<\lambda <1$.  Notice that similarly to the three dimensional case, the constant determinant of $\pi$ is constrained to be $<1$. This follows in a non-trivial way from the second constraint in (\ref{conSP2}). The solutions are again periodic with $f_\pm$ in the ranges
\be 
\cosh{f_-}<{1\over \lambda}, \quad x_1\leq e^{f_+}\leq x_2,
\ee
where $x_1$ and $x_2$ are the positive roots of
\be 
P(x)=\lambda+2x^3-3x^2.
\ee
A general solution is shown in figure \ref{fig:an}, where the three curves correspond to $f_i$ as a function of $mT$ for $\lambda =0.01$. Figure \ref{fig:an1} shows the corresponding scale factors $a_i = e^{-2f_i}$.   Observe (see figure \ref{fig:lambda}) that the solutions are characterised by two $\lambda$-dependent characteristic periods, as well as an amplitude which increases as $\lambda$ decreases.

A particular analytical solution can be found explicitly when $f_-=0$, that is for $f_1=f_2$.  In this case two of the scale factors in the metric $g_{\mu \nu}$ are identical.  
Let $x=e^{f_+}$ then energy conservation becomes
\be 
3\dot{x}^2+m^2P(x)=0.
\label{ss}
\ee
In terms of $\lambda=\cos^2\beta$, with $0\leq\beta\leq{\pi\over 2}$, the roots of $P$ are given by
\be 
x_p={1\over 2}-\cos\left({{2\beta-2p\pi\over 3}}\right),\qquad p=0,1,2.
\ee
Now introduce the new variable $\theta(T)$ through
\be 
x(T)={x_1+x_2\over 2}+{x_2-x_1\over 2}\cos\left({2\theta(T)}\right) .
\ee
Then (\ref{ss}) becomes
\be 
\dot{\theta}^2-{m^2\over 6}\left[(x_2-x_0)-(x_2-x_1)\sin^2\theta\right]=0
\ee
with implicit solution
\be 
m\sqrt{x_2-x_0\over 6}T=F(\theta,k^2),\qquad k^2={x_2-x_1\over x_2-x_0}
\ee
where 
\be 
F(\theta,k^2)=\int_{0}^\theta{d\varphi\over\sqrt{1-k^2\sin^2\varphi}}
\ee
is the Jacobi elliptic function of the first kind, and for simplicity we have taken initial conditions $\theta(0)=0$.
Replacing the roots by their values gives
\be 
m\sqrt{{\sin{2\beta+\pi\over 3}\over 2\sqrt{3}}}T=F\left(\theta, {\sin{2\beta\over 3}\over{\sin{2\beta+\pi\over 3}}}\right)
\ee
so that we finally obtain
\be 
x(T)={1\over 2}-\cos{2\beta-2\pi\over 3}+\sqrt{3}\sin{2\beta\over 3}{\rm cn}^2\left(m\sqrt{{\sin{2\beta+\pi\over 3}\over 2\sqrt{3}}}T;{\sin{2\beta\over 3}\over{\sin{2\beta+\pi\over 3}}}\right),
\ee
where ${\rm cn}F = \cos\theta$.
The period is given by 
\be 
{2\over m}\sqrt{2\sqrt{3}\over {\sin{2\beta+\pi\over 3}}} K(k^2)
\ee
where $K(k^2) = F(\pi/2,k^2)$.  The limiting case $\beta=0$ gives the linearised result $2\pi/m$ and the other limiting case is $\beta={\pi\over 2}$ (a vanishing determinant) for which the period is infinite.  Finally, for small $\beta$ one can use the series expansion of the Jacobi cn function in terms of the nome and argument (see \cite{AS}, chapter 16.23) to get the order by order perturbative solution around flat space.  In section \ref{sec:pfs} we have already found the leading order terms in the general case.

 \section{Conclusions}
 
 In this paper we have studied massive gravity in $D$-dimensions in its vielbein formulation.   This theory is parametrised by $D$ parameters $\beta_n$ which describe the different possible mass terms, and we have focused on the simplest case in which only $\beta_0$ and $\beta_1$ are non-zero.  Furthermore, we have limited ourselves to time-dependent and spatially translational invariant metrics, which in the context of Fierz-Pauli would correspond to studying plane waves, and in the context of General Relativity to Bianchi I and Kasner solutions.
 
 Our aim was two-fold. The first was to use this simplified set up to gain a clearer understanding of the degrees of freedom and constraints in massive gravity.  In general, the constraints are highly non-trivial and for this reason a Hamiltonian analysis in the general case is hard to carry out explicitly. However,  for time-dependent translational invariant systems we were able to solve the constraints exactly, and hence count the number of degrees of freedom.  As we have seen in section \ref{sec:tt} the relevant vielbeins $e_{AB}$ a priori contain $D(D+1)/2$ Lagrangian degrees of freedom,  encoded in the lapse $N(t)$, shift $n_i(t)$ and spatial components $\pi_{ij}(t)$.  The constraints however, impose that $\det(\pi)$ is a constant of motion, while the shift must identically vanish for all times.  Furthermore, the lapse $N(t)$ is then determined directly from $\pi_{ij}(t)$.  Hence phase space is simply given by the set of matrices $(\pi,\partial_t \pi)$, which are linked by two very non-linear scalar constraints: as a result phase space contains the correct $D(D-1)-2$ degrees of freedom needed to describe a massive spin 2 particle.  
  The spatial translation invariance leads to a further $(D-1)(D-2)/2$ constants of motion, encoded in the anti-symmetric matrix $\beta$ defined in (\ref{betadef}), meaning that of the $D(D-1)-2$ degrees of freedom, $(D-1)(D-2)/2 + 1$ of these are constants of motion.
  We have also shown that the constraints and equations of motion provide a well-posed initial value problem.
  
 Our second aim was to study the solutions of the system of equations and constraints.   We have shown that there are many different disconnected sectors of solutions, depending on the signature of $\pi_{ij}$. If the signature is positive definite these solutions are bounded and stable. Perturbations around flat space (see section \ref{sec:pfs}) belong to this sector of solutions. The other sectors have an effective potential which is unbounded from below and leads to runaway solutions.  Their stability in the quantum context is the subject of future work.
 
In the future it would also be interesting 
to extend this analysis to the other\footnote{for recent progress on
the study of theses cases in $D=3$ see \cite{Banados:2013fda}} $\beta_n$  where, 
as opposed to the case of non-zero $\beta_0$ and $\beta_1$ only, 
the analysis of constraints at the Lagrangian level does not 
lead to a scalar constraint which removes the ghost \cite{dmz1}.  
We expect the framework of translation invariant fields 
to greatly simplify the full analysis. 
 
 \section*{Acknowledgments}
We thank C. Deffayet and G. Zahariade for useful discussions.

\section*{Appendix: lower bound on $m$.}

\begin{figure*}
\includegraphics[width=0.85\textwidth,height=.55\textwidth]{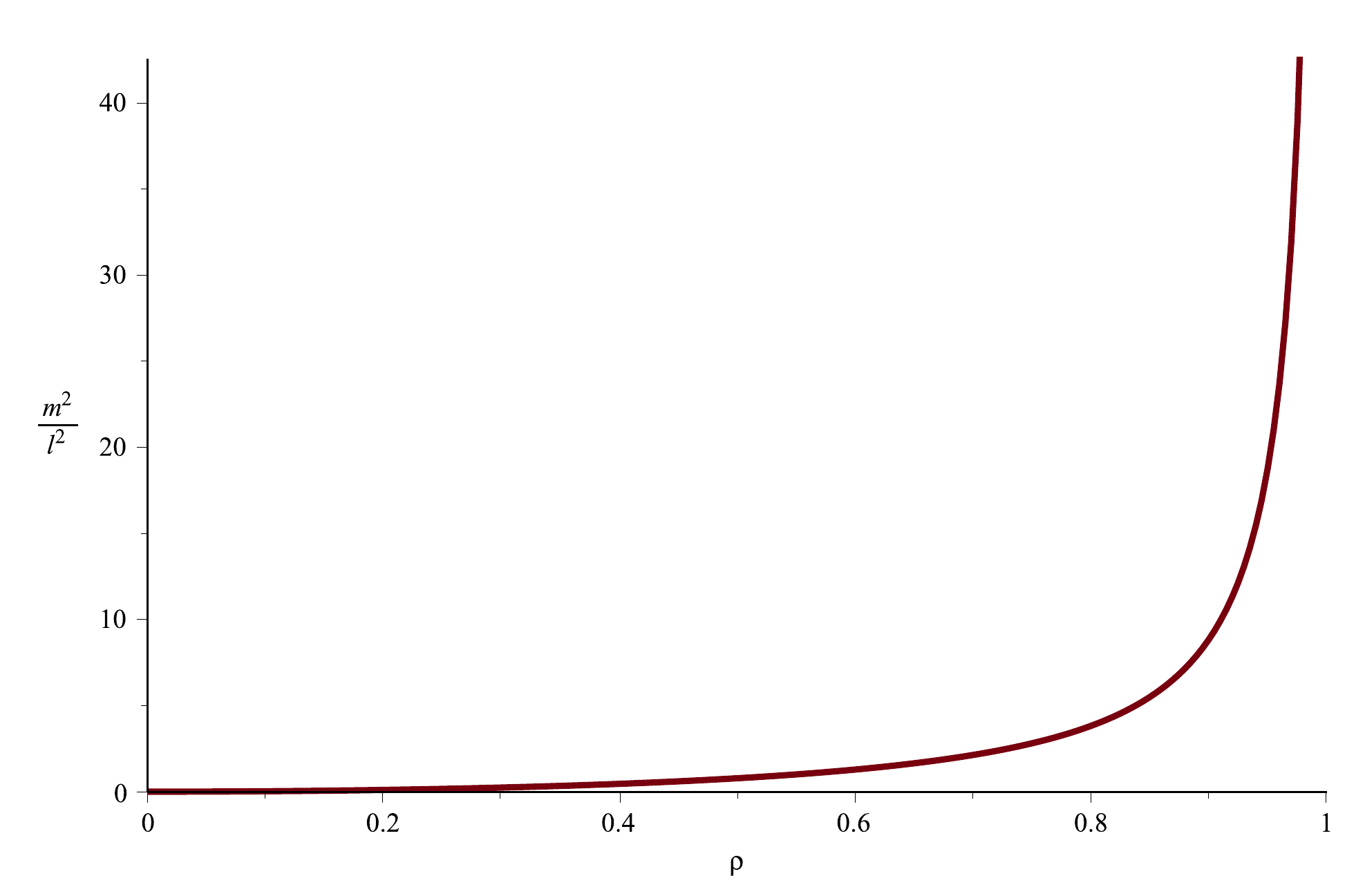}
\caption{Lower bound on $m/\ell$ as a function of $\rho$.}
\label{fig:limit}
\end{figure*}

A necessary and sufficient condition for solutions of (\ref{Vp}) to exist is that at the minimum $\xi = \bar{\xi}$, the potential $V_+(\bar{\xi}) \leq 0$.
Thus, for a given $\ell$ and $\rho$, $m=m(\ell,\rho)$ must be greater than some critical value which must satisfy
\be
V_+(\bar{\xi})=0 = \left. \frac{dV_+}{d\xi} \right|_{\bar{\xi}}  \, .
\ee
This gives two conditions with which to determine $m$ and $\bar{\xi}$, namely
\bea
\ell^2 + 2m^2\sinh^2\bar{\xi}\cosh^2\bar{\xi}(\rho \cosh\bar{\xi}-1)&=&0
\label{eq1}
\\
-\ell^2(2 \cosh^2\bar{\xi}-1) + m^2 \rho \sinh^4\bar{\xi}\cosh^3\bar{\xi} &=& 0 \, .
\label{eq2}
\eea
On eliminating the terms in $m^2\rho$ between the two equations, it follows that $2m^2 \sinh^4\bar{\xi}\cosh^2\bar{\xi}=\ell^2( 5\sinh^2\bar{\xi}+2)$ which, on substituting back into (\ref{eq2}) gives
\be
5\rho y^3 - 4y^2 - 3\rho y +2 = 0 \qquad {\rm where} \qquad  y\equiv \cosh\bar{\xi}.
\ee
This has three real roots (for $0 < \rho \leq 1$), of which only one is $>1$, and given by
\be
y(\rho)=\cosh\bar{\xi} = \frac{1}{15 \rho} \left[ 2 + \cos\left(\frac{\theta(\rho)}{3}\right)(45 \rho^2 + 16)^{1/2}\right]
\ee
where
\be
\tan \theta(\rho) = \frac{15 \rho \left( 405\rho^4 -297\rho^2 +384\right)^{1/2}}{64-405\rho^2}.
\ee
Then, using (\ref{eq1}), it follows that for a solution to exist
\be
\frac{m^2}{\ell^2} \geq \frac{5}{2y^2(y^2 + 2\rho y - 3)}
\label{finalf}
\ee 
which is plotted in figure \ref{fig:limit}.
Notice that when $\rho\rightarrow 1$, $y \rightarrow 1$ and the right hand side of (\ref{finalf}) tends to $\infty$. In the other limit $\rho=0$ we get $m/\ell \geq 0$.

\end{document}